\documentclass[twocolumn,10p]{article}
\usepackage[colorlinks=true,linkcolor=blue,citecolor=black,urlcolor=blue]{hyperref}
\usepackage[english]{babel}
\usepackage{amsmath}
\usepackage{cleveref}
\usepackage{graphicx}
\usepackage{floatrow}
\usepackage{subfig}
\usepackage{caption}
\usepackage{graphics}
\usepackage{cleveref}
\usepackage{siunitx}
\usepackage{cite}
\usepackage{amsmath}
\usepackage[section]{placeins}  

\usepackage[none]{hyphenat}
\begin{document}
\binoppenalty=10000 
\relpenalty=10000  
\title{Refined method to extract frequency-noise components of lasers by delayed self-heterodyne}

\author{N.\,Hedegaard\,Arent, 
M.\,Far\,Brusatori$^\ast$, and N.\,Volet\\ 
Aarhus University\\
Aarhus, Denmark\\
$^\ast$Email: 
{\href{mailto:mfar@ece.au.dk}{mfar@ece.au.dk}}
}

\markboth{Date \today}%
{}
\maketitle
\begin{abstract}
 \textit{An essential metric to quantify the stability of a laser is its frequency noise (FN). This metric yields information on the linewidth and on noise components which limit its usage for high precision purposes such as coherent communication.
 Its experimental determination relies on challenging optical phase measurements, for which dedicated commercial instruments have been developed.
 In contrast, this work presents a simple and cost-effective method for extracting FN features employing a delayed self-heterodyne (DSH) setup. Using delay lengths much shorter than the coherence length of the laser, the DSH trace reveals a correspondence with 
 the FN power spectral density (PSD) measured with commercial instruments. Results are found for multiple lasers,  
 with discrepancies in intense dither tone frequencies below 0.2\%. 
}
\end{abstract}
\section{Introduction}
\noindent
Narrow-linewidth lasers are critical for 
technologies such as 
LIDAR detection\,\cite{riemensberger2020massively,Hu_2014},
optical frequency combs\,\cite{raja2019packaged,Pavlov2018,suh2016microresonator},
optical frequency metrology\,\cite{frequencyMetrology,Liang2015},
frequency synthesis\,\cite{Xin2019,Spencer2018},
ultra-precise timing\,\cite{Newman2019,Shang2020},
and more fundamentally, 
the study of 
nonlinear optical phenomena
generated on a chip\,\cite{Liao2017, Volet2018,Stone2018}.
The functionalities and performance of a laser are however limited by its noise, which can be quantified by its frequency noise (FN) power spectral density (PSD).
This quantity consists of two components.
The first is intrinsic laser noise due to spontaneous emission in the gain medium.  It contributes to the noise floor of the FN PSD, which is proportional to the intrinsic or Schawlow-Townes-Henry linewidth of the laser\,\cite{Henry82,daino83,kikuchi1989effect,di2010simple,zhang2020}. 
The second results from material properties
and the system controlling the light source, such as thermoelectric coolers, and wavelength or power stabilizing loops.
This component is sometimes referred to as technical noise. Its statistical properties
give rise to \mbox{$1/f^\alpha,\, $\,($0< \alpha \leq 4$)} dependent features in the FN PSD,
along with dither tones at the operation frequency of the originating processes\,\cite{a_ziel_1988}.

The FN PSD of a laser below the relaxation resonance is composed of two major features; its intrinsic linewidth and technical noise. The latter is mostly responsible for low frequency components in the noise spectra which affect the effective linewidth of the laser\,\cite{di2010simple,mercer19911,St_phan_2005}.
%
This is of particular interest in \textit{e.g.} the field of coherent optical communication\,\cite{Pan_2013,kakkar2017laser,Zhang_2009} as lasers with lower linewidths allow for advanced modulation
formats capable of increased transmission rates. As an example, transmitting 64-QAM signals has been reported to require effective linewidths of $1.2$\,kHz \mbox{at $40$\,Gbit/s\,\cite{2008laser}}. Furthermore, the presence of dither tones in the FN PSD negatively impacts jitter tolerances on high baud rate coherent systems \cite{Zhang2020jitter}.

Methods to quantify noise are therefore necessary to identify stability issues.
%
To extract the linewidth of coherent light sources, high resolution detection techniques are required.
In the last decades, the DSH method\,\cite{okoshi1980novel} has risen to this challenge and has become a standard approach\,\cite{Horak_06} with
refinements of the 
method continuously being reported\,\cite{wang2020ultra,Ma2019,Huynh2013, Li2019, Kuse2018}.
An alternative approach relies on measuring the FN PSD of the light source. Various methods have been investigated that typically involve post processing of the measured signal\,\cite{Kikuchi2012, Brajato_20, Camatel2008,Turner2002}. In addition, several manufacturers offer dedicated equipment\,\cite{HFinesse,sycatus,$OE$waves},
allowing for a straightforward analysis of its noise components, in addition to relative intensity noise (RIN) measurements.
%

This work proposes that features of the FN PSD of a laser can be observed in its DSH trace by exploring its highly coherent regime. This is in contrast to the traditional incoherent regime used to estimate laser instrinsic linewidth.
In the following sections, a study of five commercial laser is presented, which evidences a direct correspondence between the DSH trace and the FN PSD without the need of data post-processing.
Results are validated by comparison with a dedicated commercial instrument, making the proposed method a cost-effective solution for analyzing laser stability. 

\section{Experimental setup}
\label{sec:setup}
\begin{figure}[h]
	\centering
	\includegraphics[width=0.75\linewidth]{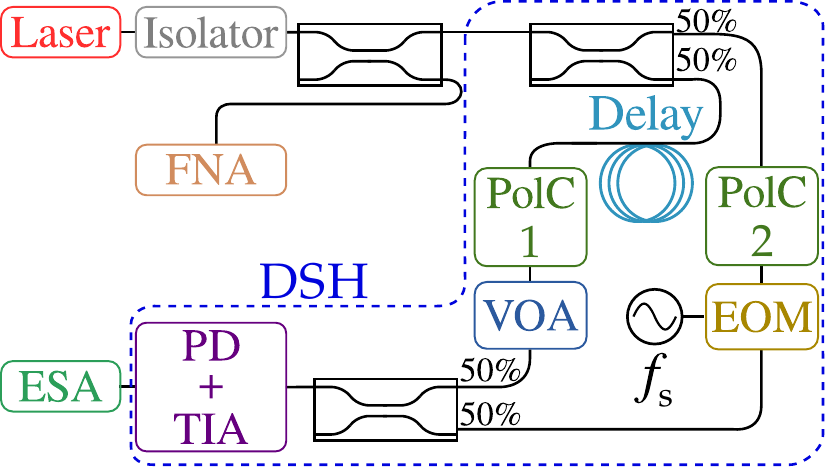}
	\caption{Schematic of the setup used in this work, where the DSH arrangement is delimited by a blue box.
	}
	\label{fig:DSH_setup}
\end{figure}

The experimental setup 
is displayed in Fig.\,\ref{fig:DSH_setup}.
The laser passes through an optical isolator followed by a splitter. One output is sent to a frequency-noise analyzer (FNA; HighFinesse, LWA-1k-1550) and the other into a DSH arrangement. 
%
%
The 
latter splits the input light to form an unbalanced Mach-Zehnder interferometer (UMZI). One branch includes a fiber delay of length $L_\mathrm{d}$, a polarization controller (PolC, Thorlabs) and a variable optical attenuator (VOA). The second branch contains an electro-optic modulator (EOM, iXblue MPZ-LN-01) driven at a frequency $f_\mathrm{s}$ 
using a signal generator (Siglent SDG6022X). A second PolC is used to maximize the intensity of the side-bands.
The beat signal from the recombined light is detected by an amplified photo-detector (PD) and recorded by an electrical spectrum analyzer (ESA). 

\begin{figure}[h]
	\centering
	\includegraphics[width=0.8\linewidth]{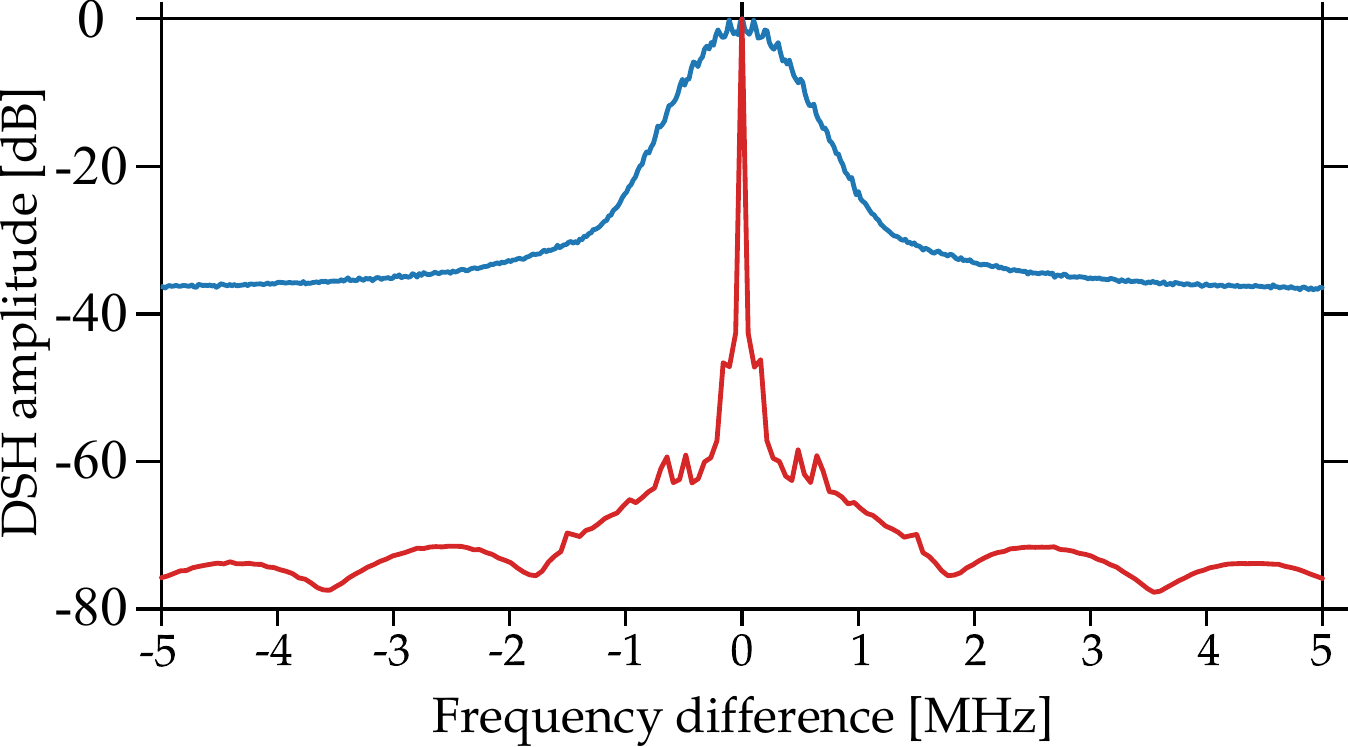}
	\caption{DSH traces for a Pure Photonics PPCL550 laser,
	in the coherent (in red) and incoherent (in blue) regimes.}
	\label{fig:DSH_traces_exp}
\end{figure}

The DSH trace depends on the ratio between the fiber delay length, $L_\mathrm{d}$, and the laser coherence length, $L_{\mathrm{coh}}$\,\cite{Richter1986}. The latter can be expressed as\cite{Henry82}:
%
%
    \begin{align}
        L_{\mathrm{coh}} &\approx
        \frac{c}{n_\mathrm{g}} \frac{1}{\pi \Delta f},
        \label{eq:coherence_length}
    \end{align}
%
where $\Delta f$ is the intrinsic linewidth of the laser, $n_\mathrm{g}$ is the effective group index (for single-mode fiber at $1550$\,nm wavelength, $n_\mathrm{g} = 1.48$) and $c$ is the speed of light in vacuum. 
%
%
The DSH method traditionally employs a delay $L_\mathrm{d} > L_{\mathrm{coh}}$,
which translates to incoherent signals detected on the PD.
In this case, the DSH trace has
a Voigt profile\,\cite{armstrong1967spectrum}, from which the lasers intrinsic linewidth is estimated as the HWHM of the Lorentzian part.
%
%
%
%
As an example, Fig.\,\ref{fig:DSH_traces_exp} shows the two distinct regimes of DSH measurements for a Pure Photonics PPCL550 laser. The horizontal axis corresponds to the frequency difference between the ESA trace and $f_{\mathrm{s}}$, and the vertical scale is normalized to the maximum power.
The red curve in Fig.\,\ref{fig:DSH_traces_exp} illustrates the case for $L_\mathrm{d} < L_{\mathrm{coh}}$,
where the phases of the incident fields are coherent. This results in an oscillating pattern with a period given by $1/t_\mathrm{d}$, where $t_\mathrm{d}$ is the time delay due to the fiber delay, in addition to a delta peak at the carrier frequency, $f_\mathrm{s}$\,\cite{huang2017precise}.
%
%
The blue curve corresponds to a DSH trace obtained when the setup operated in the incoherent regime.
The intrinsic linewidth is extracted from the Lorentzian component of the observed Voigt profile.
%
%
%
%
%

In contrast, 
this work uses 
highly coherent DSH measurements, \textit{i.e.} $L_\mathrm{d} \ll L_{\mathrm{coh}}$. In practice, this involves choosing a fiber delay short enough so as not to observe the oscillations previously described, and long enough so that some temporal difference is withheld between the two arms in the UMZI. If the latter criteria is not met, only a delta peak at $f_\mathrm{s}$ is observed in the DSH trace.
The following sections detail the obtained results, which show that in this regime,
features
of the FN PSD can be observed. This is validated by comparison with data obtained with an FNA.

\section{Results and discussion}
Results are divided in two sections, corresponding to both configurations of ESA and amplified PD used in this work. 
%
All the FN PSDs shown are an average of 12 - 25 traces.

\vspace{-1mm}

\subsection*{\centering ESA: Siglent - PD: Thorlabs}
The results shown in this section are measured using an amplified PD (Thorlabs PDA05CF2), and an ESA (Siglent SSA3021X) whose signal is averaged 20 times.
%
%
%
\begin{figure}[h]
\sidesubfloat[]{\includegraphics[width=0.4\linewidth]
{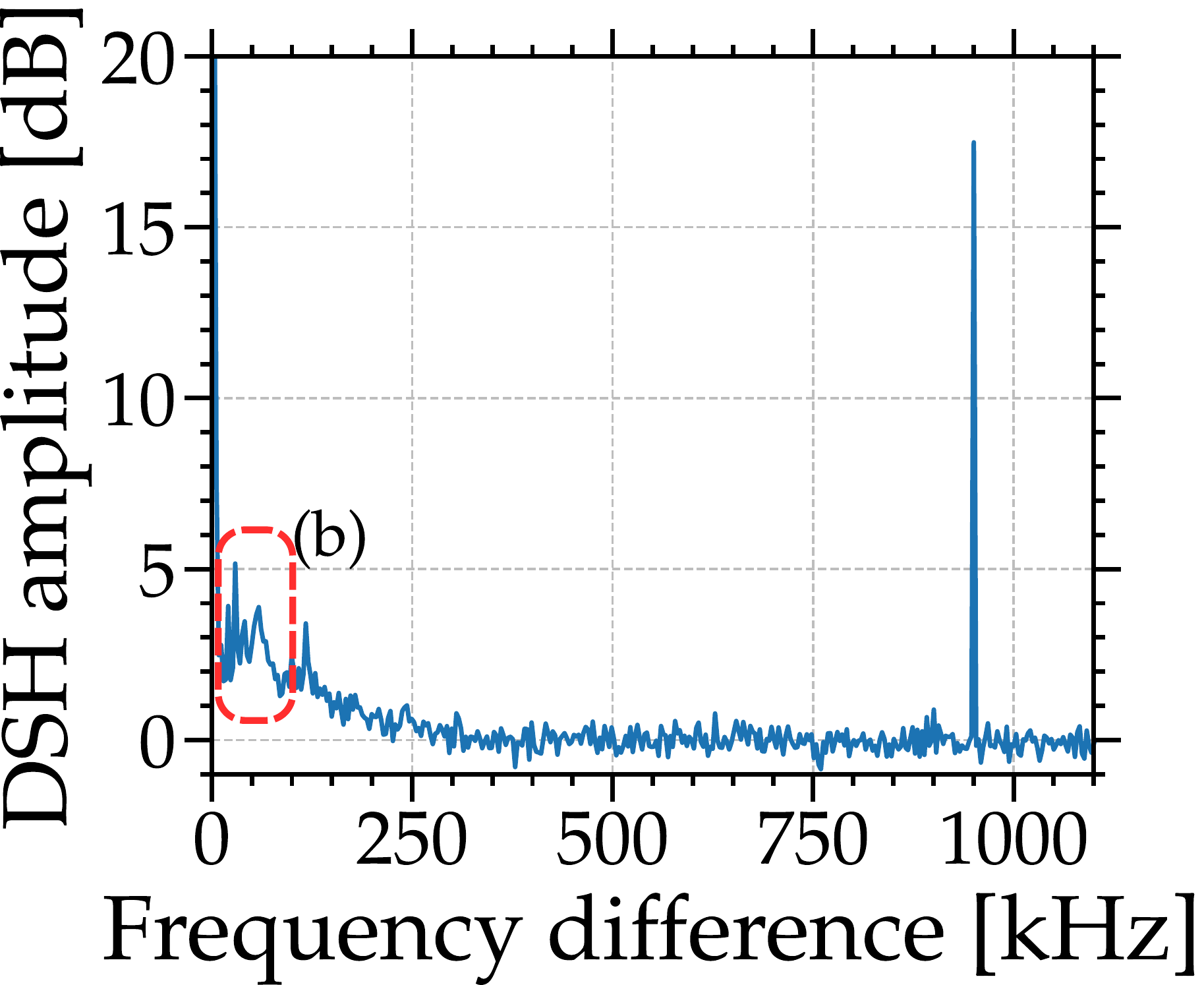}}
\hspace{3mm}
\sidesubfloat[]{\includegraphics[width=0.4\linewidth]
{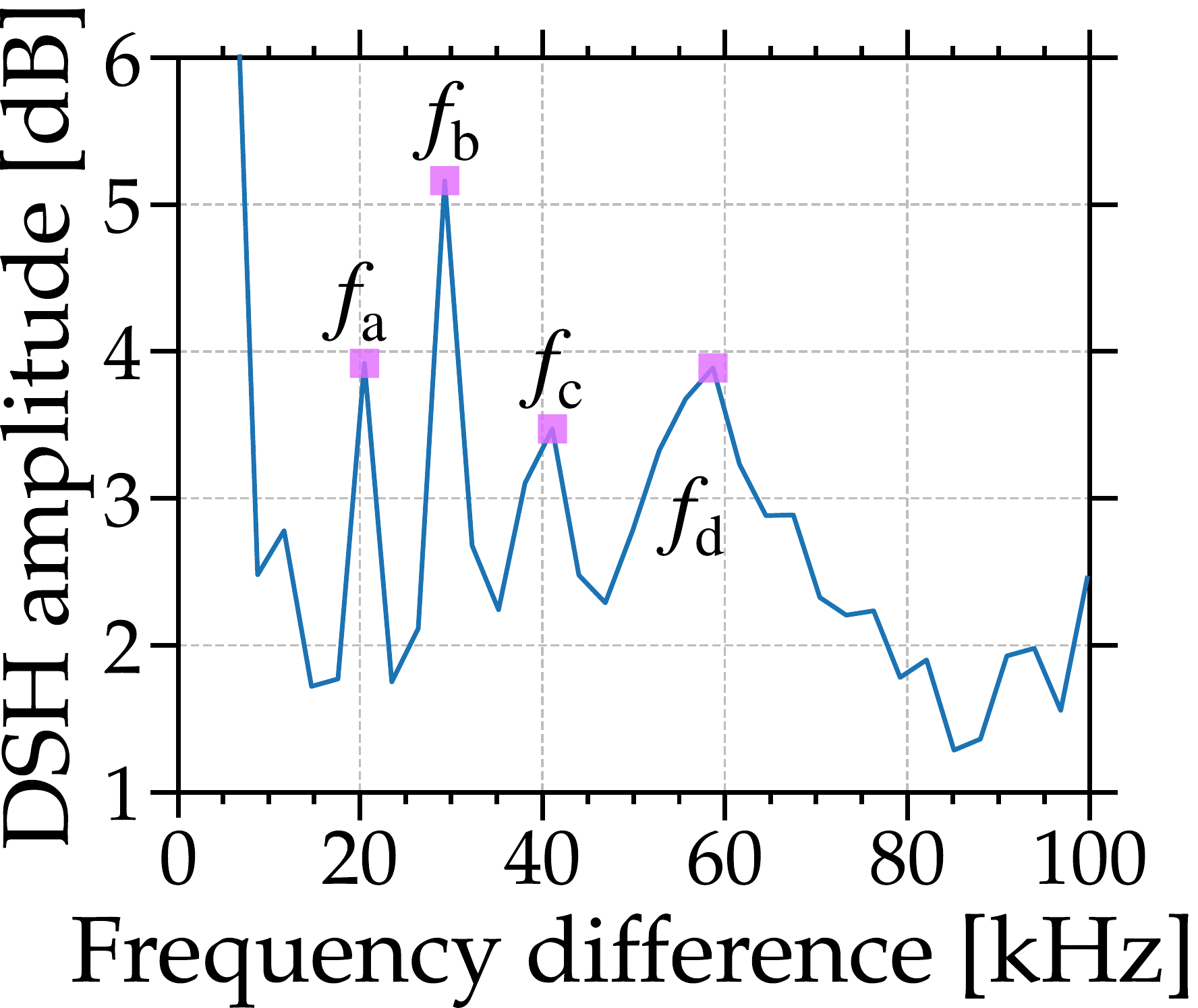}}
\vspace{3mm}
\\
\sidesubfloat[]{\includegraphics[width=0.4\linewidth]
{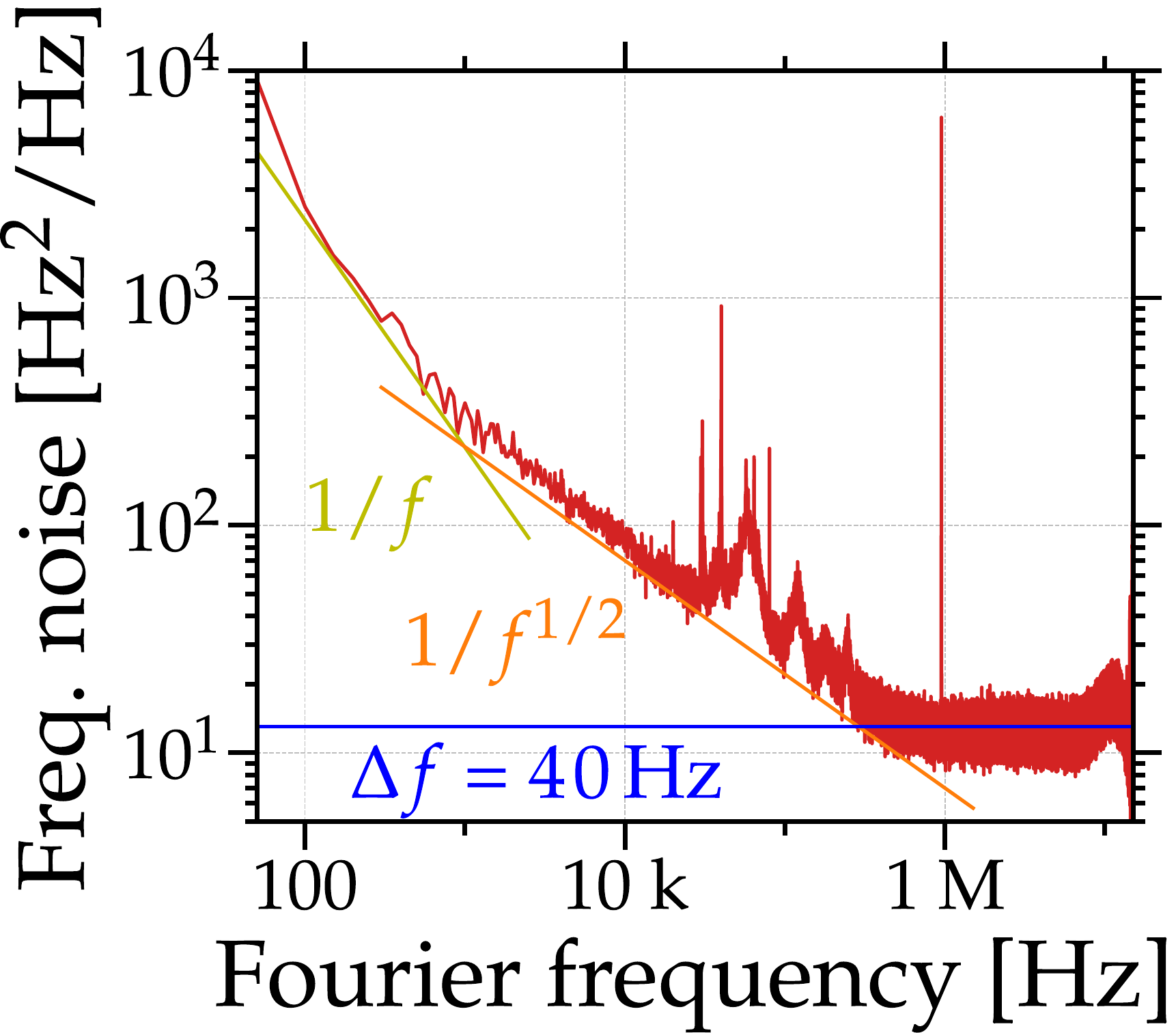}}
\hspace{1mm}
\sidesubfloat[]{\includegraphics[width=0.4\linewidth]
{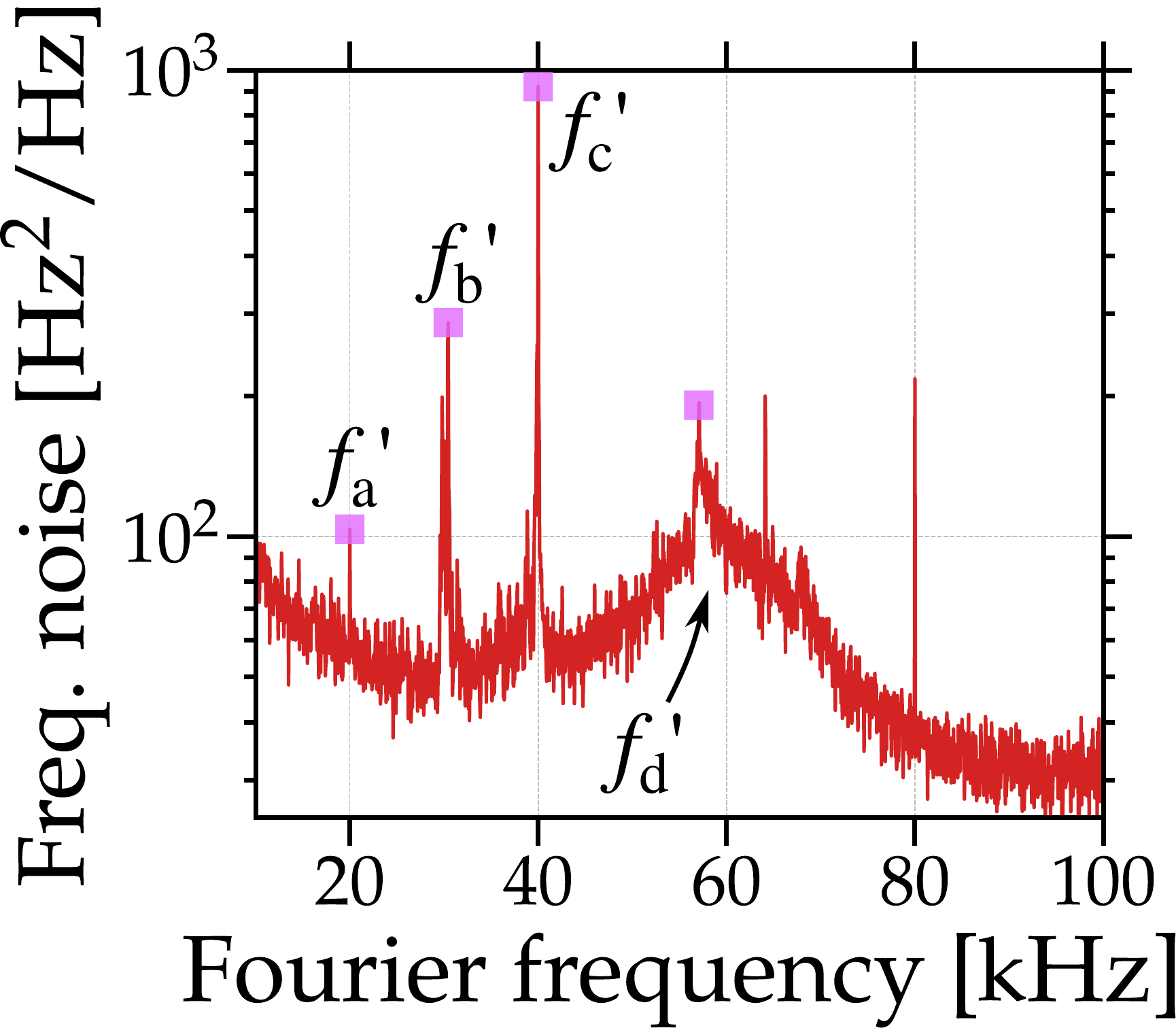}}
\caption{\textbf{(a)} 
DSH trace for the E15 Koheras laser using 100\,m fiber delay. Area delimited by a red rectangle in \textbf{(a)} is magnified in \textbf{(b)}.  \textbf{(c)} Corresponding FN PSD trace \textbf{(d)} with
a magnification in the same range as for \textbf{(b)}.}
	\label{fig:DSH_E15}
\end{figure}
The modulation frequency is set to $f_\mathrm{s}=120$\,MHz and acts as the central frequency of the DSH trace. This value is chosen to stay beyond the low frequency noise from external
sources and to keep within the $150$-MHz-bandwidth of the PD.

%
%
\mbox{Fig.\,\ref{fig:DSH_E15}\,(a)-(b)} shows the DSH trace obtained for a NKT Koheras E15 fiber laser using $100$\,m of fiber delay. The vertical scale is normalized to the noise floor.
From \mbox{Fig.\,\ref{fig:DSH_E15}\,(c)} an intrinsic linewidth of $\approx40$\,Hz ($L_{\mathrm{coh}} \approx 1600$\,km) is extracted. This value is in agreement with the data sheet of the laser\;($<0.1$ kHz) \cite{NKTP_E15_specs}. The delay length used in this experiment 
is thus well 
within the coherent regime of the DSH method. Note that the intrinsic linewidth of the E15 is reported to be below the minimum value of the FNA ($\sim$300 Hz). The value obtained for the intrinsic linewidth from the measured FN PSD is an estimate, however it is possible to assume it is below $\sim314$\,Hz ($L_{\mathrm{coh}} = 184$\,km). Even for this upper bound, the fiber delay chosen is short enough to be within the highly coherent regime.


As seen in \mbox{Fig.\,\ref{fig:DSH_E15}\,(b)}, four amplitude peaks are discerned, centered at $f_\mathrm{a} = 20$\,kHz, $f_\mathrm{b} = 29$\,kHz, $f_\mathrm{c} = 41$\,kHz and $f_\mathrm{d} = 59$\,kHz. Despite the limiting resolution ($3$\,kHz), a less intense peak can be observed at  $12$\,kHz. This peak is likely located at $10$\,kHz, and as such is the fundamental tone of the subsequent peaks. 
%
A high amplitude peak at $950$\,kHz is also observed in Fig.\,\ref{fig:DSH_E15}\,(a).
A comparison of these results with the FN PSD are shown in Fig.\,\ref{fig:DSH_E15}\,(c)-(d), extracted with the FNA mentioned in Sec.\,\ref{sec:setup}. The higher resolution of the instrument allows to detect multiple strong dither tones. By focusing on the frequency values corresponding to the previously mentioned DSH peaks, peaks are detected at $f'_\mathrm{a} = 20.00$\,kHz, $f'_\mathrm{b} =30.10$\,kHz, $f'_\mathrm{c} =40.00$\,kHz and $f'_\mathrm{d} =57.10$\,kHz. These values have an uncertainty of $50$\,Hz which presents no significant differences with the frequencies found using the DSH method. While the peak at 12\,kHz is not observed, it is likely hidden by the $1 / f$ components of the electrical noise present in the trace. 
The tones appearing at integer values of 10\,kHz likely originate due to an oscillating process present in the laser or in the surrounding instrumentation. Methods to mitigate its influence on the light source could be thus developed using the information provided by this method. The noise tone observed at $950$\,kHz is also measured for a different version of the laser from the same manufacturer. An explanation is that this tone originates from an internal, generic temperature stabilization loop since these are known to operate at this range of frequencies\,\cite{a_ziel_1988}.

From this analysis it is possible to conclude that 
the proposed DSH technique shows both intense narrow features of the FN PSD (\textit{e.g}. $f_\mathrm{a}$, $f_\mathrm{b}$ and $f_\mathrm{c}$), and less intense but wide features, likely of acoustic origin, such as that observed at $f_\mathrm{d}$.

\vspace{-1mm}

\subsection*{\centering ESA: Rohde \& Schwarz - PD: Finisar}

To increase the frequency accuracy of the dither tones found using the DSH method, additional measurements are carried out using detection with decreased resolution. This allows for a more suitable comparison with the FN PSD signal obtained.
Traces are measured with an ESA R\&S FSW50, and additionally
100 traces are captured to obtain a mean value and standard deviation of the amplitude of the spectra. 
The beat signal is detected using a PD with 50-GHz bandwidth (Finisar XPDV2320R). A higher modulation frequency of $f_\mathrm{s}=190$\,MHz is chosen, as the ESA used has a lower noise floor at higher frequencies.
Additionally, the fiber delay length is chosen at $2$\,m, which is within the coherent regime for all lasers under study.
This section shows results obtained for 4 different lasers.
%




\subsubsection{EXFO laser}
\begin{figure}[!b]
\sidesubfloat[]{\includegraphics[width=0.9\linewidth]
{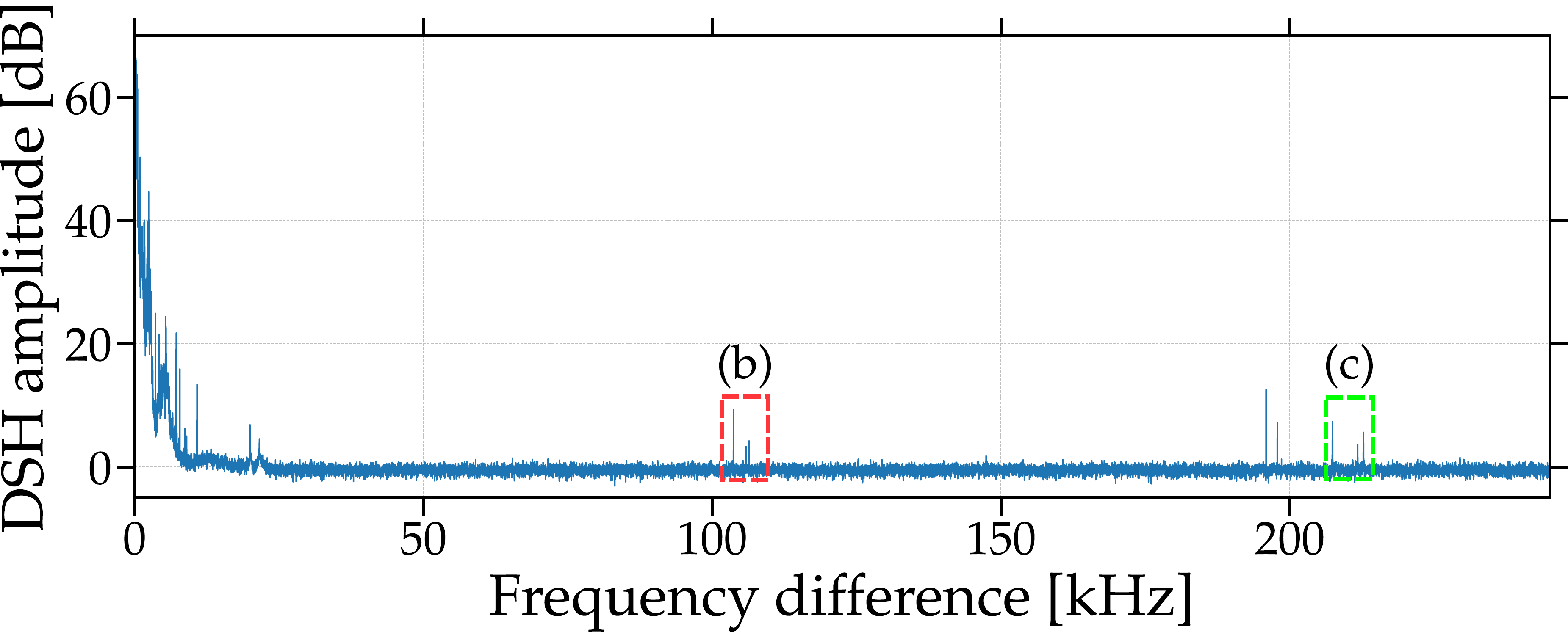}}\vspace{3mm}
\\
\sidesubfloat[]{\includegraphics[width=0.43\linewidth]{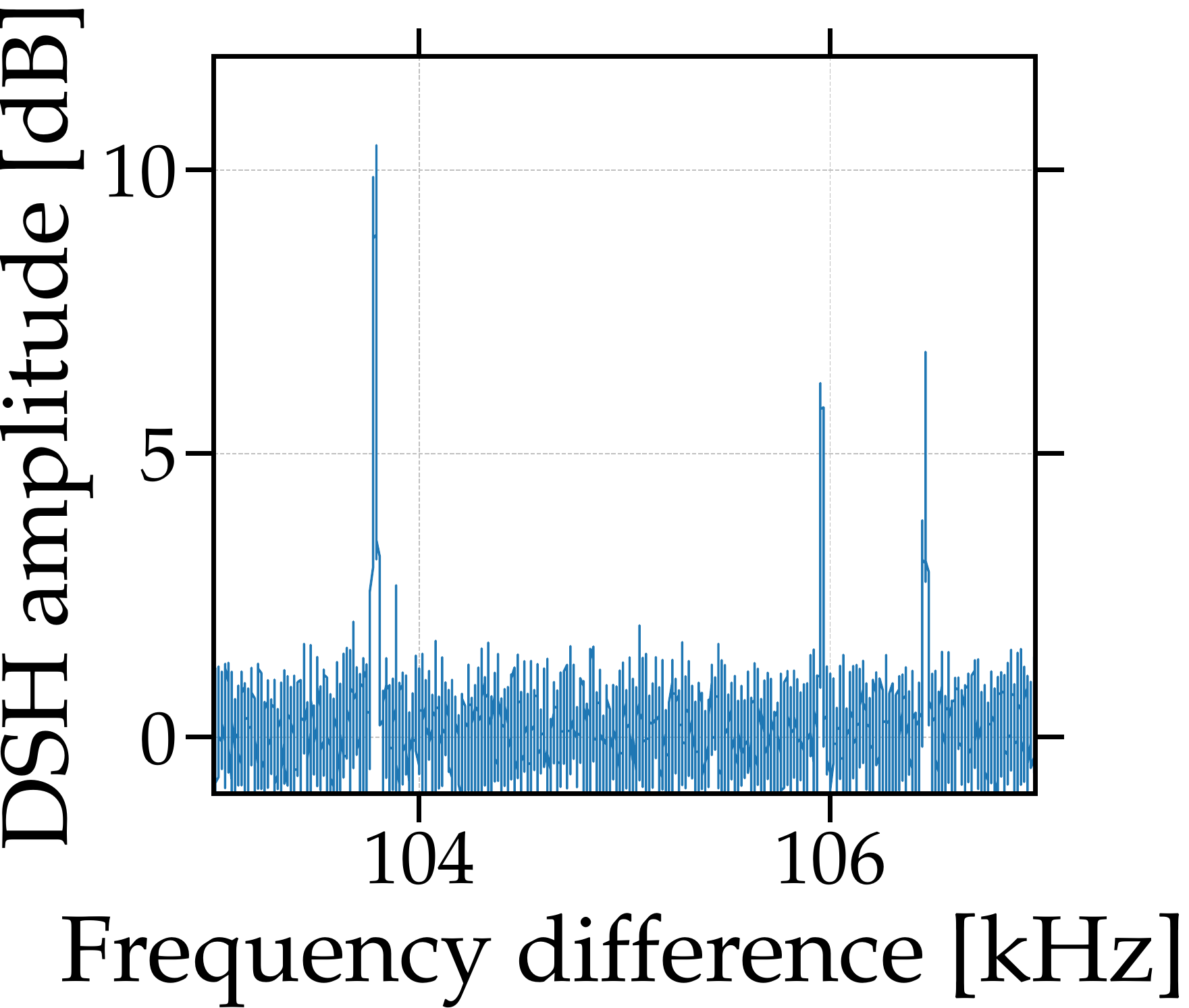}} \vspace{3mm}
\sidesubfloat[]{\includegraphics[width=0.43\linewidth]{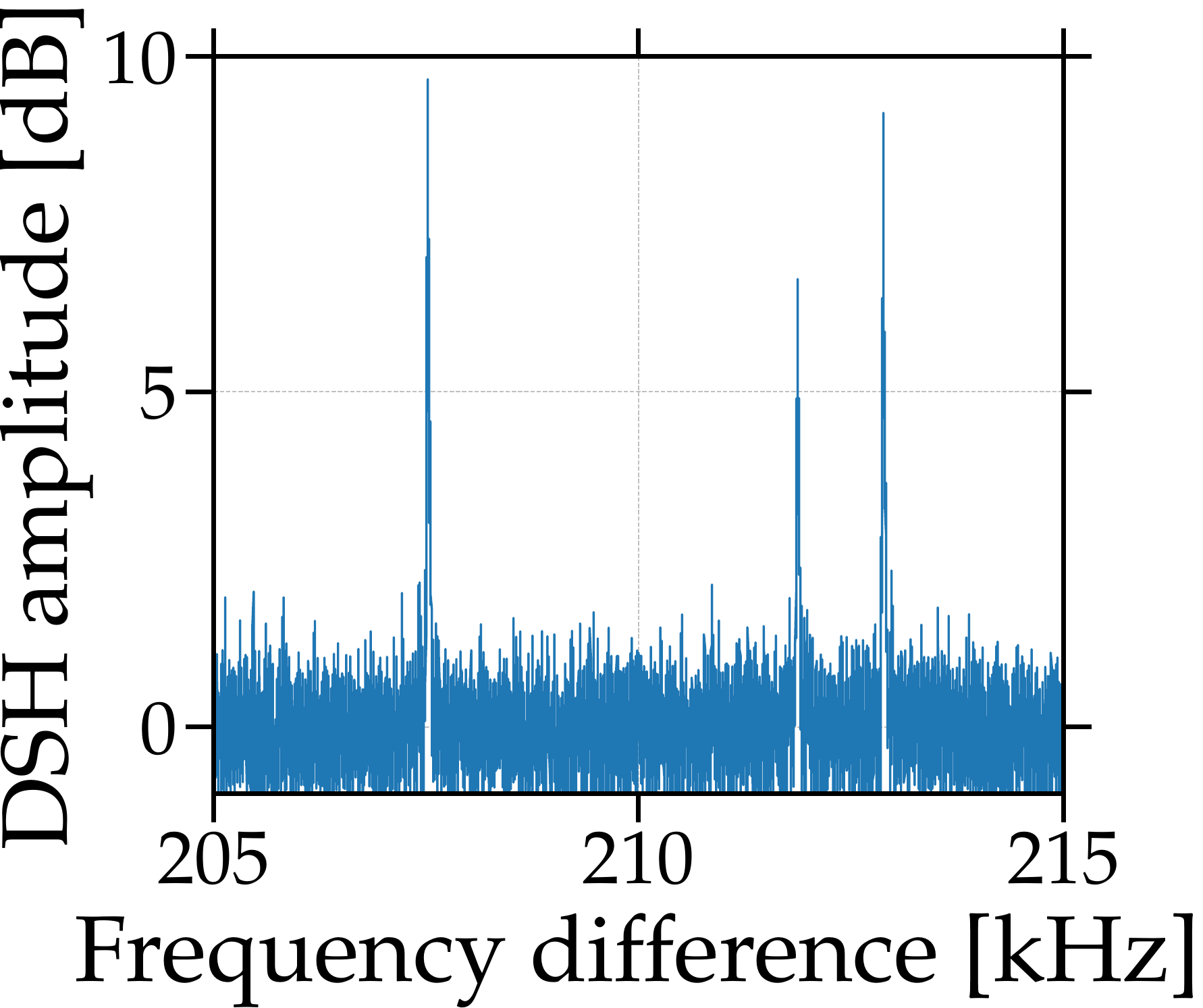}} \hspace{1.5mm}
\\
\sidesubfloat[]{\includegraphics[width=0.9\linewidth]
{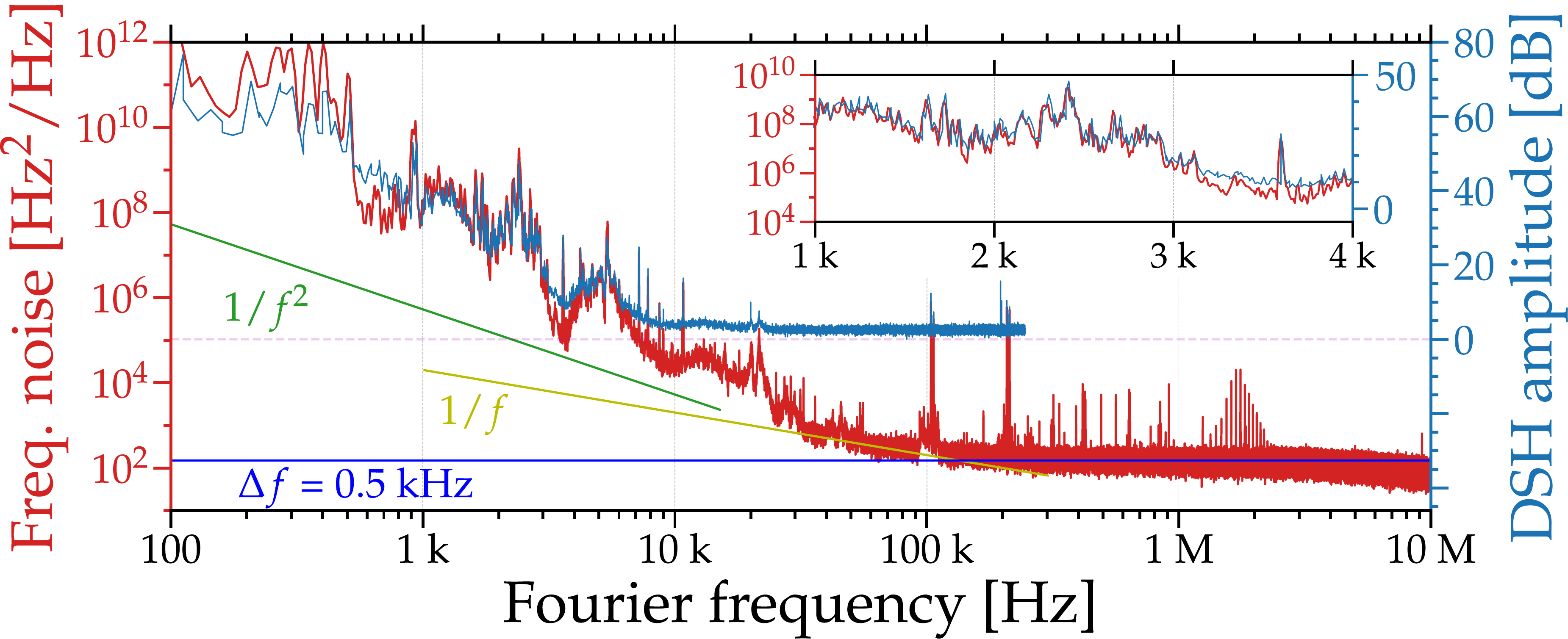}}
\caption{ \textbf{(a)}  DSH trace measured for the EXFO laser using a delay length of $2$\,m. Area delimited by a red and a green rectangles are magnified in parts \textbf{(b)} and \textbf{(c)}, respectively.
\textbf{(d)} FN PSD  (in red) and DSH trace (in blue) of the EXFO laser.}
	\label{fig:EXFODSH}
\end{figure}

The DSH trace measured for an EXFO T100S-HP laser is shown in  Fig.\,\ref{fig:EXFODSH}, obtained for an injection current of 200 mA ($\sim$12 mW of output power).
Fig.\,\ref{fig:EXFODSH}\,(a) displays the full trace, and Fig.\,\ref{fig:EXFODSH}\,(b) and Fig.\,\ref{fig:EXFODSH}\,(c) magnify sections where strong dither tones are present.
Fig.\,\ref{fig:EXFODSH}\,(b) shows three distinct peaks at frequencies $103.68$\,kHz, $105.95$\,kHz, $106.46$\,kHz. Furthermore,  Fig.\,\ref{fig:EXFODSH}\,(c) shows three prominent peaks at frequencies $207.79$\,kHz, $211.87$\,kHz, $212.88$\,kHz, which correspond to the harmonics of the frequency peaks shown in Fig.\,\ref{fig:EXFODSH}\,(b). The uncertainty in these values is of $20$\,Hz.
Fig.\,\ref{fig:EXFODSH}\,(d) shows a direct comparison between the FN PSD, acquired with the linewidth analyzer previously metioned, and the DSH trace measured for the EXFO laser. The figure also includes an estimation of its $1/f^2$ and $1/f$ noise components as well as its intrinsic linewidth. In addition, a zoom of the frequencies between 1 to 4 kHz is included in the inset, where a strong correspondence between both traces is observed. Furthermore, focusing on the ranges of the DSH trace shown in Fig.\,\ref{fig:EXFODSH}, prominent peaks are found in the FN PSD at the frequencies $103.71$\,kHz, $105.88$\,kHz, $106.40$\,kHz, 
$207.49$\,kHz, $211.77$ and $212.80$\,kHz. These values have an uncertainty of $10$\,Hz, and differ in less than $0.004\%$ with  those seen in the DSH trace. 
%
%

It is worth mentioning that not all peaks are present in both traces. 
For example, the linewidth analyzer captures higher harmonics of the peaks shown in Fig.\,\ref{fig:EXFODSH}\,(b) and Fig.\,\ref{fig:EXFODSH}\,(c), which are not seen in the DSH trace likely due to the noise floor level of the ESA used. Another possible explanation is that the dither tones measured by the PD are unrelated to the laser, emerging from the instruments used in the setup or from external sources.
%




\subsubsection{ANDO laser}
\begin{figure}[h]
\sidesubfloat[]{\includegraphics[width=0.4\linewidth]{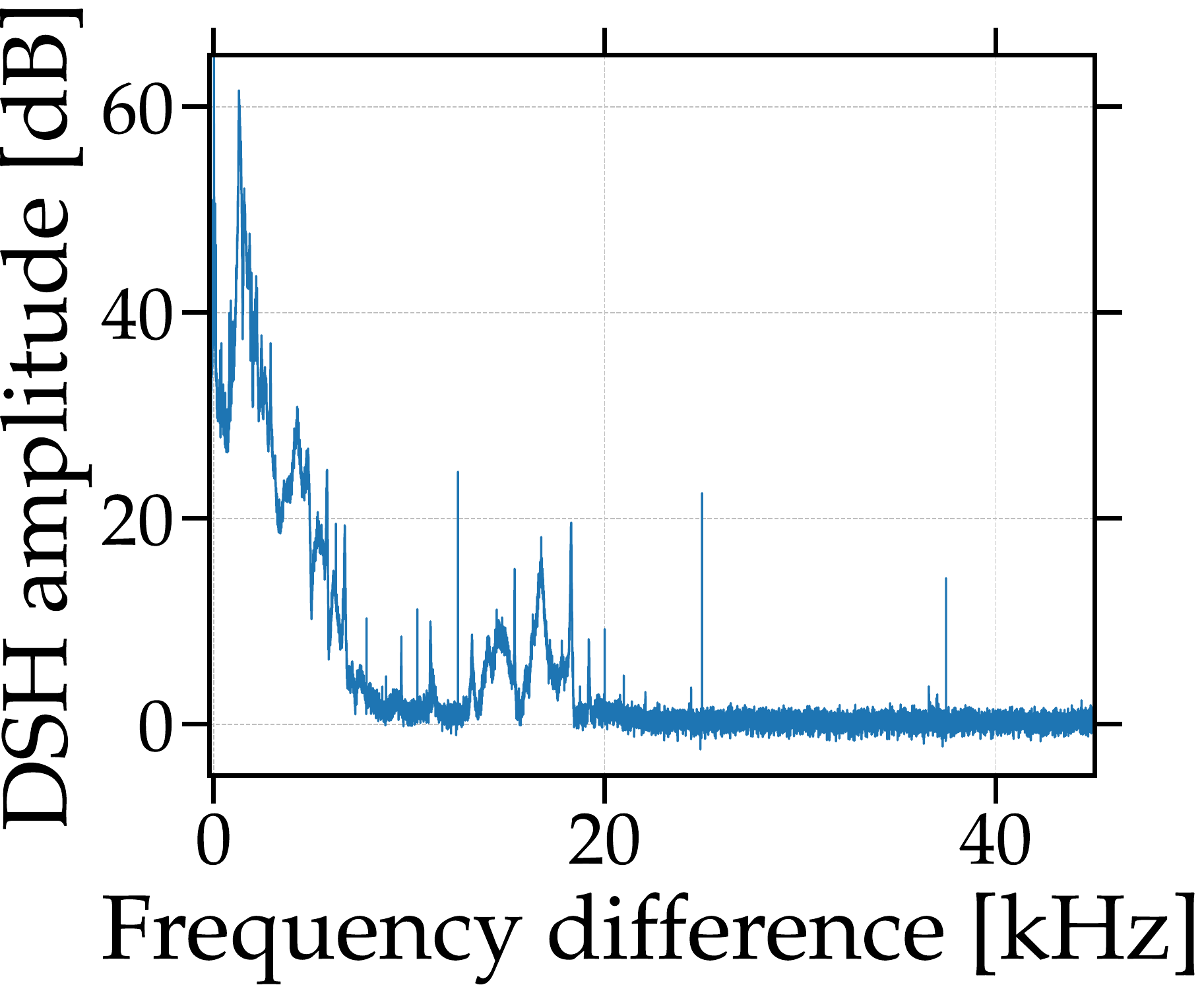}}
\hspace{2mm}
\sidesubfloat[]{\includegraphics[width=0.4\linewidth]{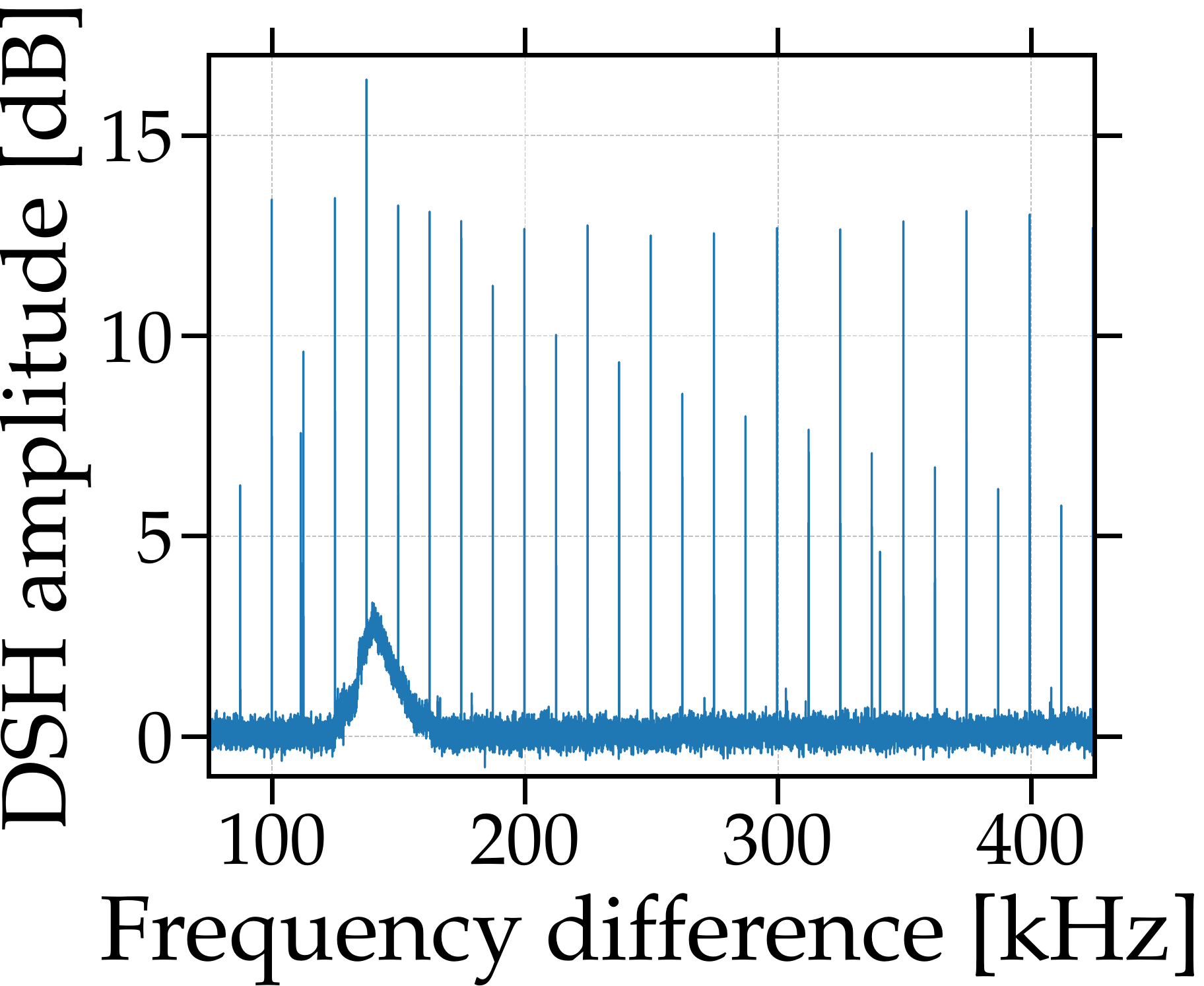}}
\vspace{2mm}
\\
\sidesubfloat[]{
\includegraphics[width=0.9\linewidth]{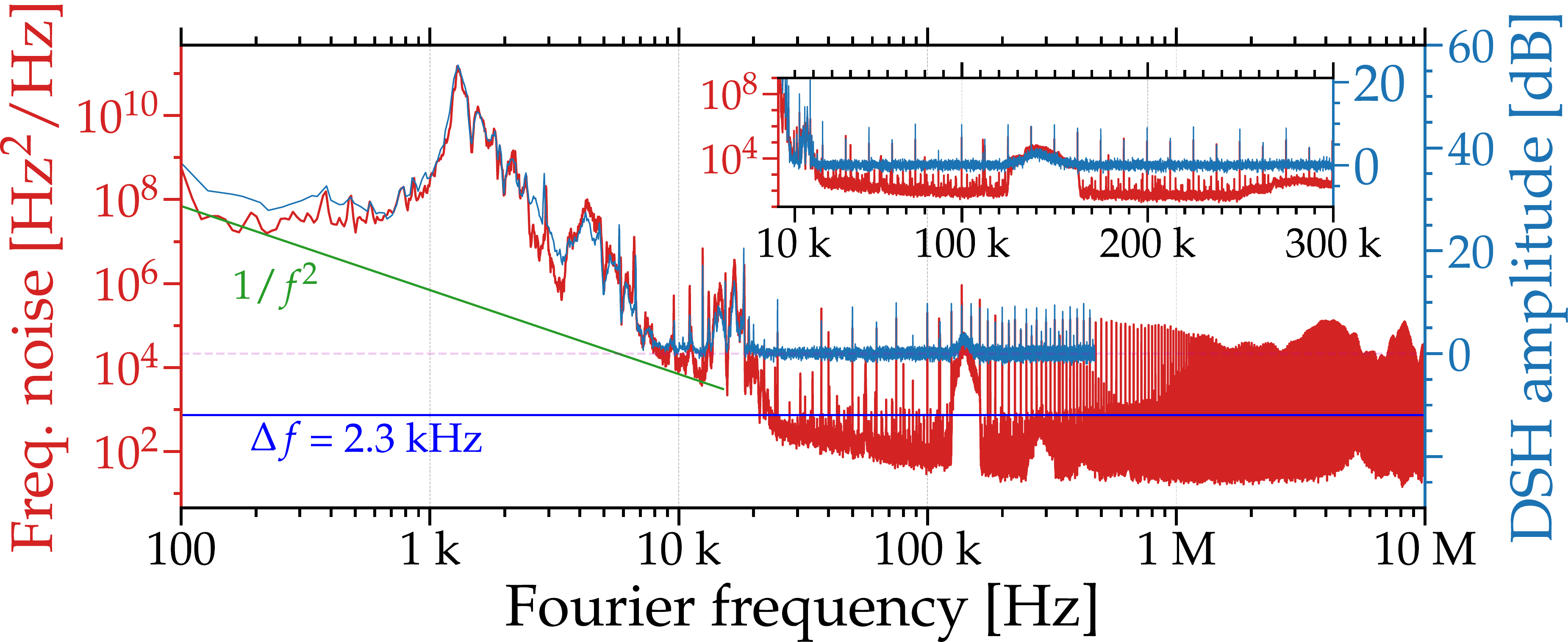}}
\caption{\textbf{(a-b)} DSH trace measured for the ANDO laser. \textbf{(c)} FN PSD  (in red) and DSH trace (in blue) for the ANDO laser.}
	\label{fig:andoDSH}
\end{figure}

Fig.\,\ref{fig:andoDSH}\,(a-b) shows the DSH trace obtained for an ANDO AQ-4321D laser operated at $6.3$\,mW.
Fig.\,\ref{fig:andoDSH}\,(a) shows the low frequency range, where both wide and narrow features are present, and Fig.\,\ref{fig:andoDSH}\,(b) focuses on an area where a myriad of evenly spaced dither tones are present. The spacing between these peaks is $12.48$\,kHz with an uncertainty of $20$\,Hz.
Fig.\,\ref{fig:andoDSH}\,(c) shows the FN PSD for the same laser with an overlay of the corresponding DSH trace.
%
The inset shows a range where both the DSH trace and the FN PSD display a high number of harmonics. The spacing of said peaks in the FN PSD is of $12.48$\,kHz, with a resolution uncertainty of 10Hz. This shows no significant differences between the spacing measured with both methods. Furthermore, a peak at this spacing frequency can be observed in both traces, in Fig.\,\ref{fig:andoDSH}\,(b) and\,\ref{fig:andoDSH}\,(c), more precisely at $12.50$\,kHz for the former and $12.48$\,kHz for the latter, and once again both values are within instrumental uncertainties. These tones are likely originated in the driving mechanisms of this laser.
Note once again that certain low intensity peaks can only be seen in the FN PSD signal, which is likely due to either the resolution in the ESA or its noise floor.
%
%
%
%
%
%
%
Qualitatively, the overall shape of the traces shows strong correspondence, presenting similar structural features at the same frequencies for the full span of the figure, particularly in the range below $30$\,kHz. 

%

%
%




\subsubsection{Agilent laser}
\begin{figure}[!b]
\sidesubfloat[]{\includegraphics[width=0.4\linewidth]{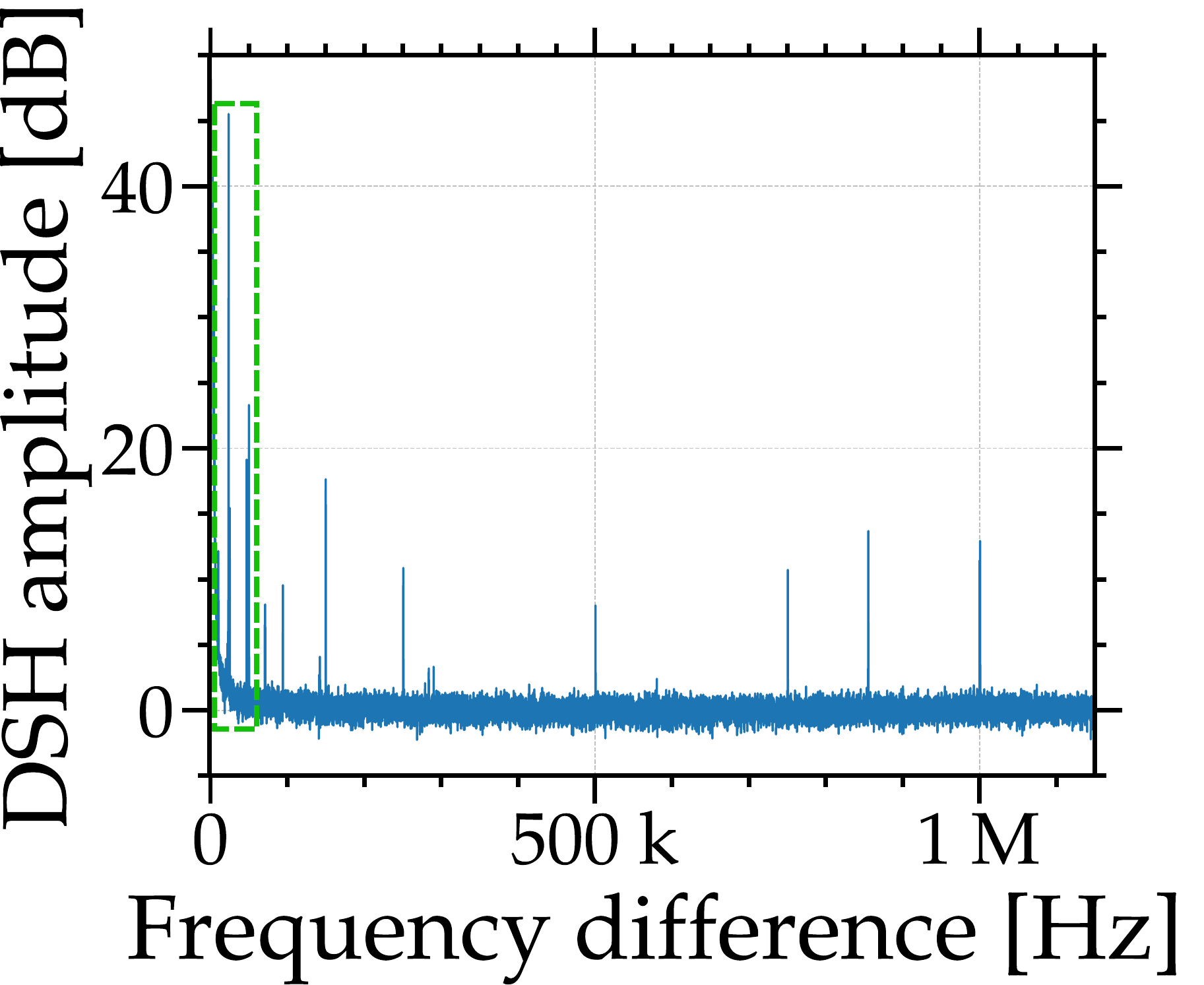}}
\hspace{1mm}
\sidesubfloat[]{\includegraphics[width=0.4\linewidth]{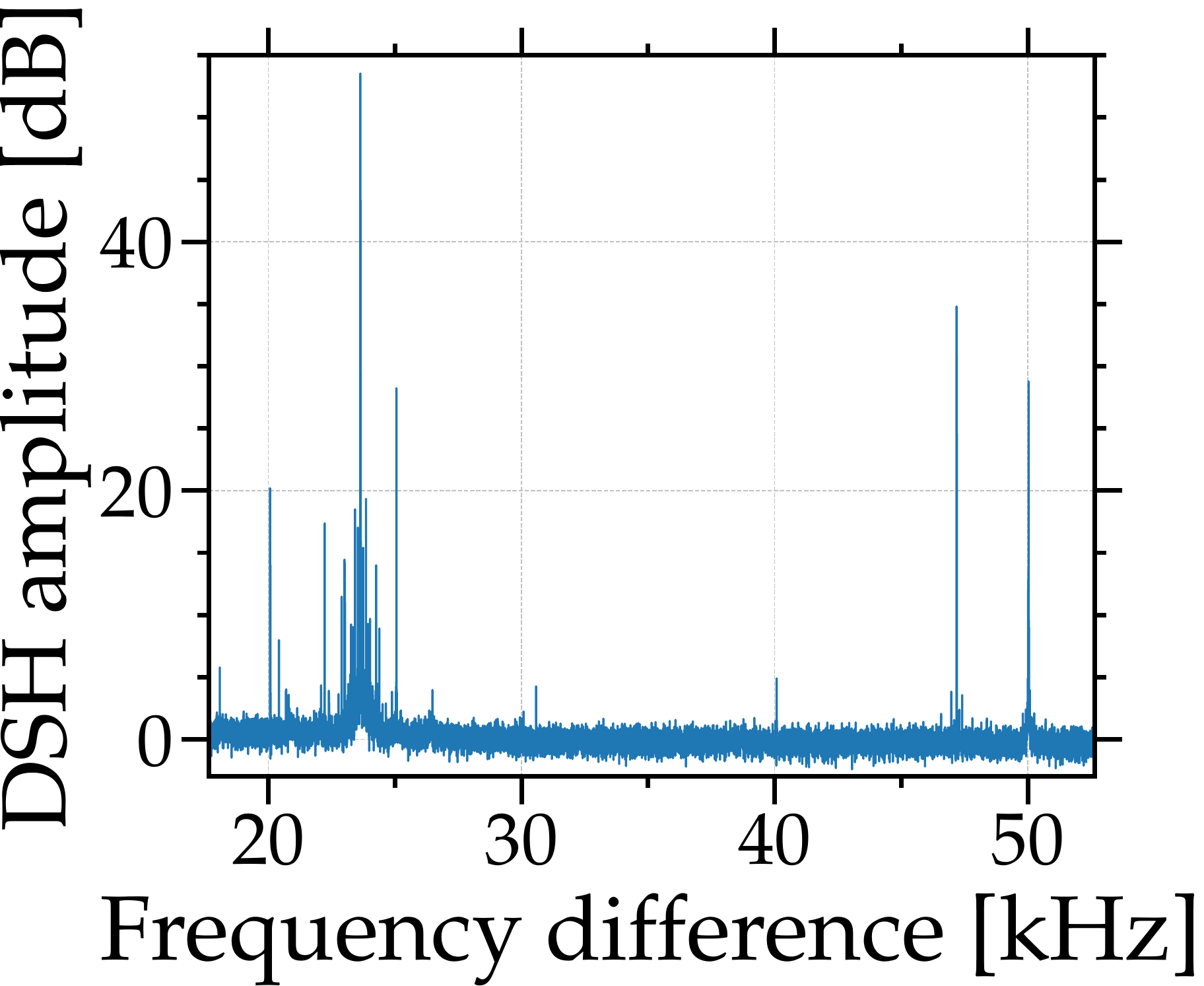}}
\vspace{2mm}
\\
\sidesubfloat[]{\includegraphics[width=0.9\linewidth]{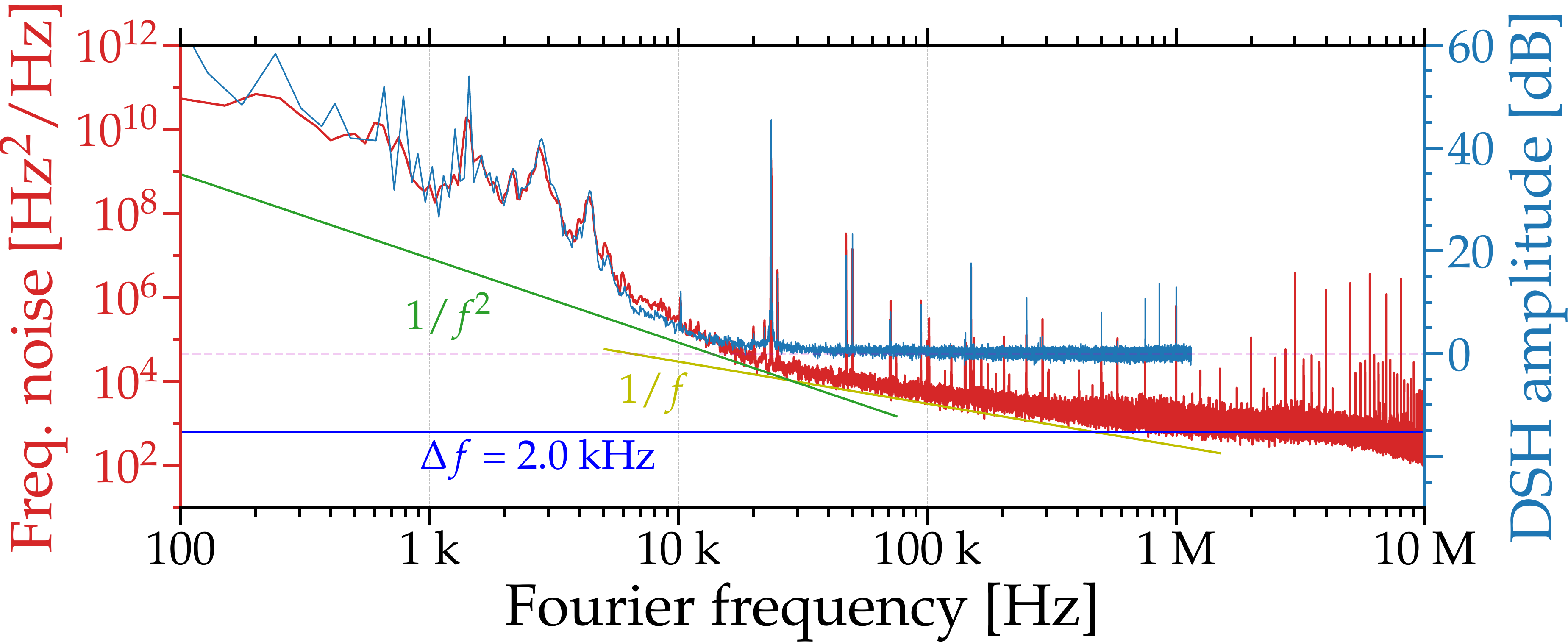}}
\caption{\textbf{(a)} DSH trace measured for the Agilent laser. Area delimited by a green rectangle in \textbf{(a)} is magnified in \textbf{(b)}. \mbox{\textbf{(c)} FN PSD  (in red) and DSH trace} (in blue).}
	\label{fig:agilentDSH}
\end{figure}

The full DSH trace measured for an Agilent 8164B laser driven at $13$\,mW is shown in Fig.\,\ref{fig:agilentDSH}\,(a). 
A magnified section shown in Fig.\,\ref{fig:agilentDSH}\,(b) where both wide and narrow features are present.
%
The comparison between the FN PSD  of this laser and its DSH trace is seen in Fig.\,\ref{fig:agilentDSH}\,(c). A peak analysis is performed in the ranges shown. This yields 21 peaks found in both traces at the same frequencies within the measurement uncertainties. An additional 7 peaks are found at frequencies differing in less than 0.2\%. 
Qualitatively, and as in previous cases, both the DSH signal and the FN PSD follow a similar amplitude pattern for both wide and narrow features.
%

%




\subsubsection{Pure Photonics PPCL550 laser}

\begin{figure}[!b]
\sidesubfloat[]{\includegraphics[width=0.4\linewidth]{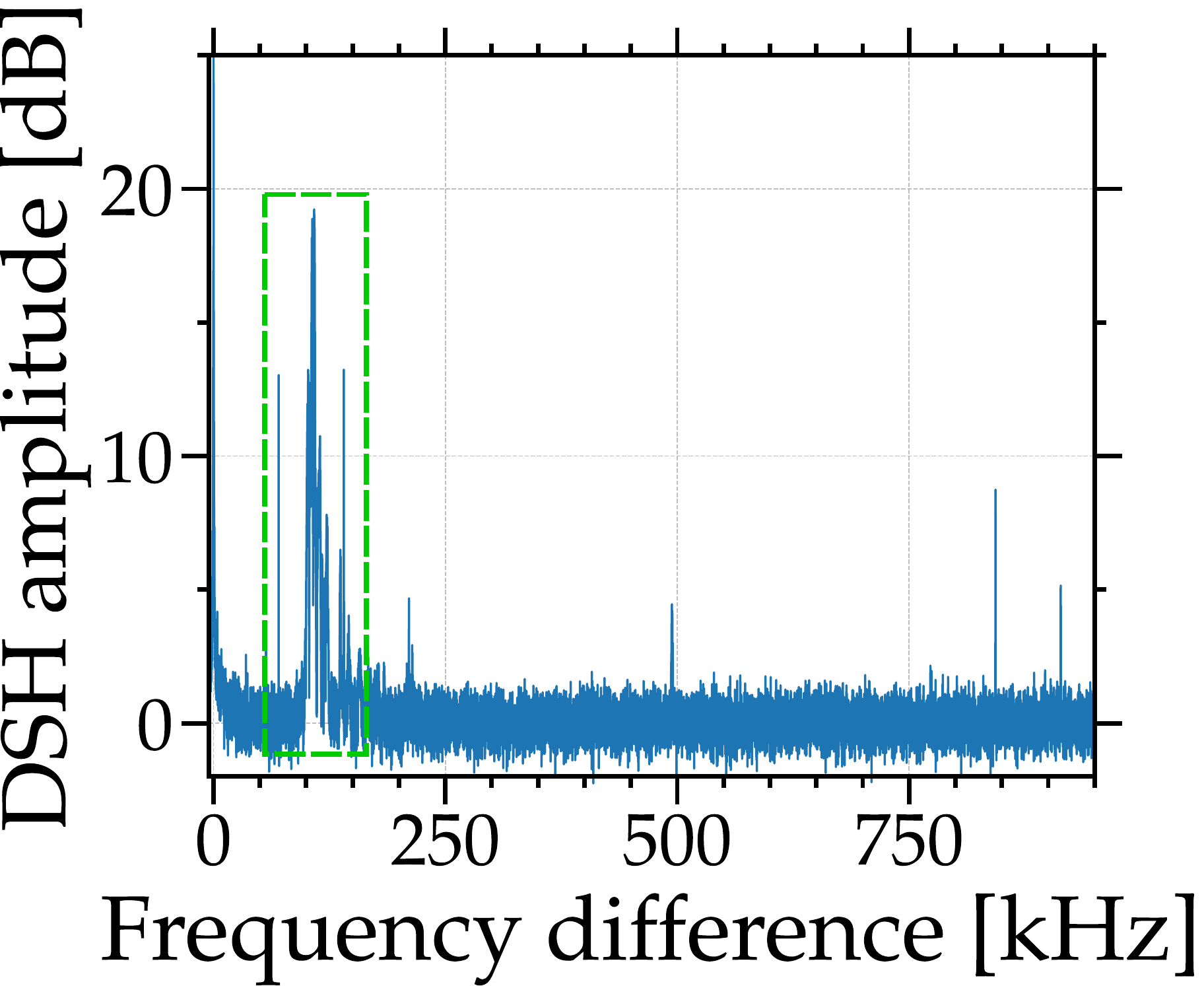}}
\hspace{1mm}
\sidesubfloat[]{\includegraphics[width=0.4\linewidth]{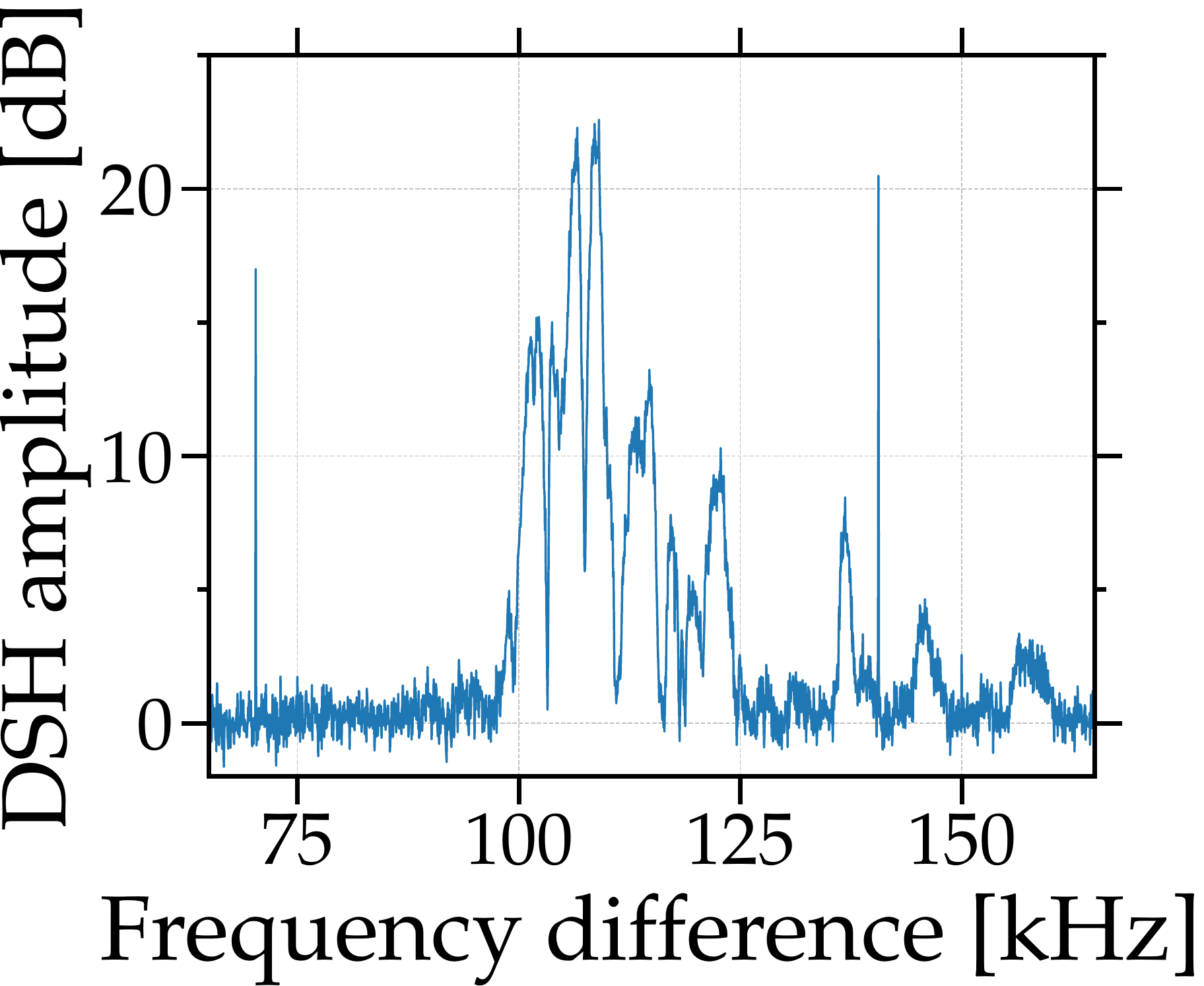}}
\vspace{3mm}
\\
\sidesubfloat[]{\includegraphics[width=0.9\linewidth]{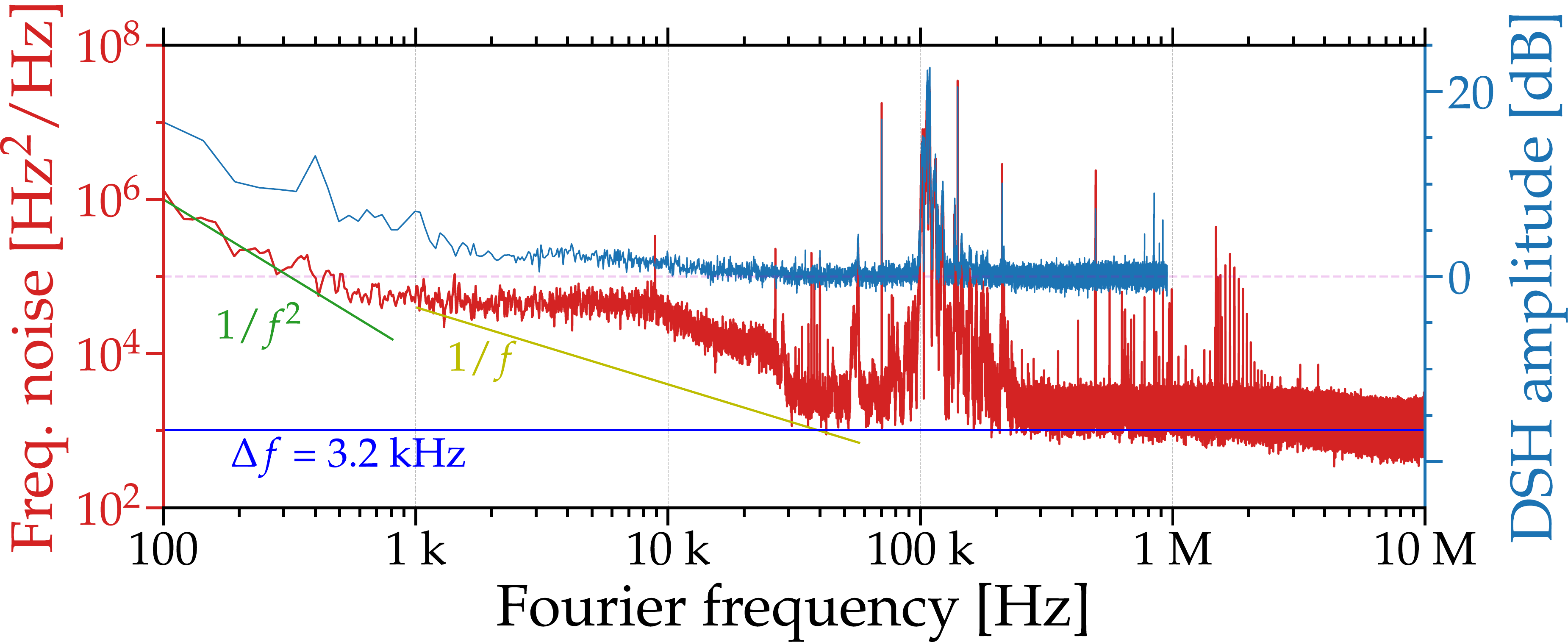}}
\vspace{1mm} 
\\
\sidesubfloat[]{\includegraphics[width=0.9\linewidth]{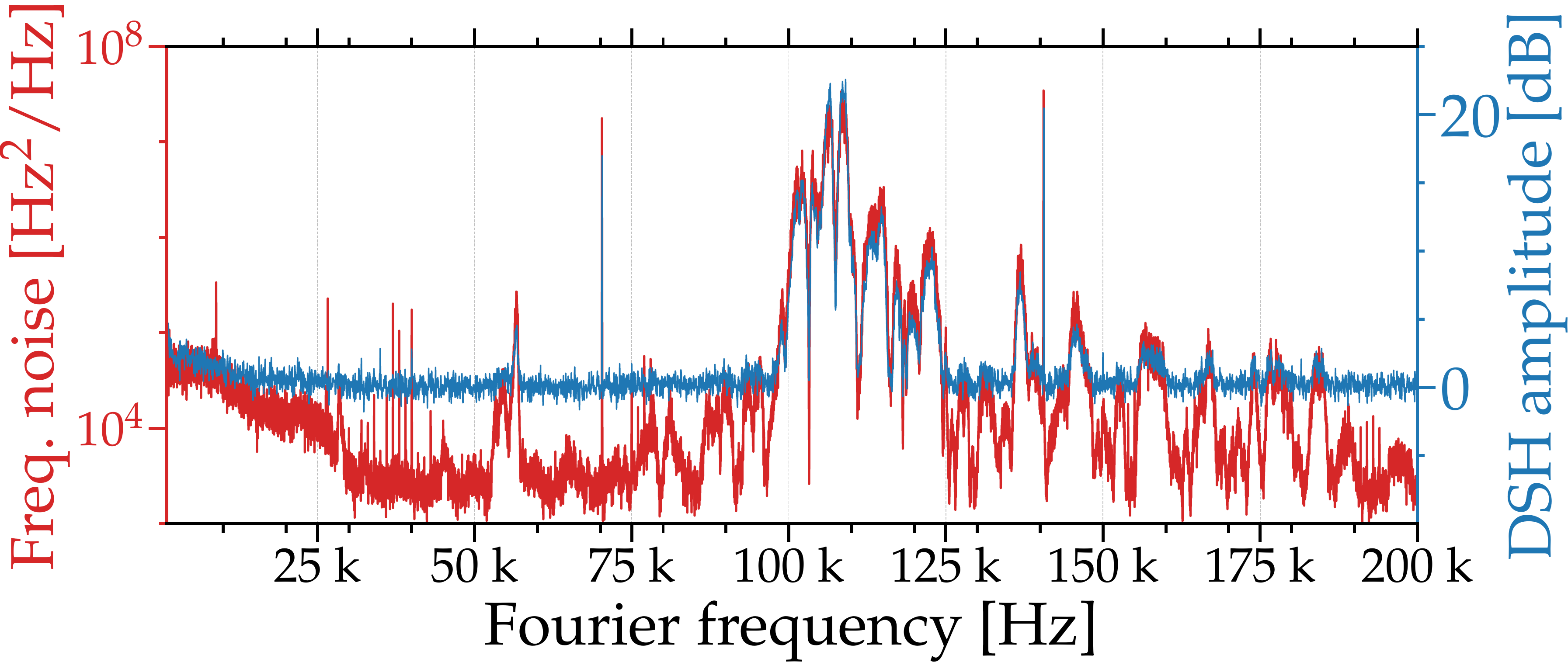}}
\caption{\textbf{(a-b)} DSH trace measured for the Pure Photonics laser.
Area delimited by a green rectangle in (a) is magnified in \textbf{(b)}.
\textbf{(c)} FN PSD  (in red) and DSH trace (in blue) for  the Pure Photonics laser. Area delimited by a green rectangle in \textbf{(c)} is magnified in \textbf{(d)}.
}
	\label{fig:PPDSH}
\end{figure}

The DSH trace obtained for a Pure Photonics PPCL550 is shown in Fig.\,\ref{fig:PPDSH}\,(a), and an intense wide feature is magnified in Fig.\,\ref{fig:PPDSH}\,(b). Finally, both the DSH trace and FN PSD are shown in Fig.\,\ref{fig:PPDSH}\,(c), with Fig.\,\ref{fig:PPDSH}\,(d) showing a magnification around the feature previously shown in Fig.\,\ref{fig:PPDSH}\,(b).
Both traces have relative amplitudes that follow the same shape, and high intensity peaks are observed at matching frequencies.
A peak analysis is performed in the ranges shown, from which 22 peaks are found in both traces which show no significant differences in frequency. Additionally, 21 peaks are found at frequencies differing in less than 0.09\% between both methods.

As a final note,
Figs. \ref{fig:EXFODSH}\,(d), \ref{fig:andoDSH}\,(c), \ref{fig:agilentDSH}\,(c) and \ref{fig:PPDSH}\,(c) show DSH traces with a noise floor at approximately $-155$\,dBm. If this noise floor were to be lowered, it is expected that the DSH method would be able to resolve noise components with lower amplitude. This could lead to an approximation of the $1/f$ and $1/f^2$ contributions, as well as allowing to distinguish dither tones that are concealed by either those contributions or the noise floor.

\section{Conclusion}
The analysis presented in this paper shows that the signal obtained with a DSH setup operated in the coherent regime contains information that allows to pinpoint different noise contributions in lasers.
In contrast to the FN PSD measurement, this is obtained without the need of data postprocessing. This is validated by a comparison with FN PSD measurements performed with an FNA. A peak analysis of various intense dither tones yields excellent match between both methods, with discrepancies below 0.2\% in their frequency, with at least half of them showing no significant differences. In addition, a detailed comparison between the traces obtained with these methods shows a clear correspondence in the relative amplitude and overall shape between the signals.
This method still has limitations, as the traces obtained from the FNA show spectral features not seen in the DSH trace. These are, however, likely concealed by the noise floor of the instrumentation used, and are expected to be revealed should the noise floor be lowered.
To conclude, the approach explored in this work increases the versatility of the DSH method by using it to extract the FN PSD components of a laser. Results show good overlap with commercial instruments, particularly in the low frequency range, which critically influences the effective linewidth of lasers. 
As this method does not rely on dedicated instruments, it results in a cost-effective solution to assess laser stability. 



\section{Acknowledgements}
The authors would like to thank the Ion Trap Group from the Department of Physics at Aarhus University and especially Frederik Skifter Tribler for facilitating the experiments performed on the E15/X15 Koheras laser.
This work is supported by Independent Research Fund Denmark (DFF).

\bibliography{refs_for_journal.bib}

\end{document}